\documentclass[11pt, oneside]{article}      
\usepackage{geometry}                		
\usepackage[parfill]{parskip}    		         
\usepackage{graphicx}				 
\usepackage{amssymb}
\usepackage[T1]{fontenc}
\usepackage{authblk}
\usepackage[utf8]{inputenc}

\title{Co-citations in context: disciplinary heterogeneity is relevant\footnote{Accepted for publication in Quantitative Science Studies.}}

\author[1]{James Bradley}
\author[2]{Sitaram Devarakonda}
\author[2]{Avon Davey}
\author[2]{Dmitriy Korobskiy}
\author[2]{Siyu Liu}
\author[2]{Djamil Lakhdar-Hamina}
\author[3]{Tandy Warnow}
\author[2]{George Chacko\thanks{netelabs@nete.com}}

\affil[1]{Raymond Mason School of Business, Coll. of William \& Mary, Williamsburg, VA}
\affil[2]{Netelabs, NET ESolutions Corporation, McLean, VA}
\affil[3]{Department of Computer Science, Univ. of Illinois, Urbana-Champaign, IL}

\begin{document}
\maketitle
\newpage

\begin{abstract}
Citation analysis of the scientific literature has been used to study and define disciplinary boundaries, to trace the dissemination of knowledge, and to estimate impact. Co-citation, the frequency with which pairs of publications are cited, provides insight into how documents relate to each other and across fields. Co-citation analysis has been used to characterize combinations of prior work as conventional or innovative and to derive features of highly cited publications. Given the organization of science into disciplines, a key question is the sensitivity of such analyses to frame of reference. Our study examines this question using  semantically-themed citation networks. We observe that trends reported to be true across the scientific literature do not hold for focused citation networks, and we conclude that inferring novelty using co-citation analysis and random graph models benefits from disciplinary context.
\end{abstract}

\section{Introduction} Citation and network analysis of  scientific literature reveals information on semantic relationships between publications, collaboration between scientists, and the practice of citation itself \cite{garfield_citation_1955,de_solla_price_networks_1965,newman_structure_2001,Shi:2010:CHI:1816123.1816131,patience_pmid28560354,stigler_1994}. Co-citation, the frequency with which two documents are cited together in other documents, provides additional insights, including the identification of semantically related documents, fields, specializations, and new ideas in science \cite{small_co-citation_1973, marshakova-shaikevich_co-citation_1973,boyack_co-citation_2010, 10.3389/frma.2018.00020,wang_bias_2017}.  

In a novel approach, Uzzi and colleagues \cite{uzzi_atypical_2013} used co-citation analysis to characterize a subset of highly cited articles with respect to both novel and conventional combinations of prior research. The frequency with which references were co-cited  in 17.9 million articles and their cited references from the Web of Science (WoS) was calculated and expressed as journal pair frequencies (observed co-citation frequencies). Expected co-citation values were generated using Monte Carlo simulations under a random graph model. Observed frequencies were then normalized (shifted and scaled) to averaged expected values from ten randomized networks and termed as \emph{z-scores}. Consequently, every article was associated with multiple z-scores corresponding to co-cited journal pairs in its references. For each article, positional statistics of z-scores were calculated and evaluated to set thresholds for a binary classification of conventionality using the median z-score of an article, and novelty using the tenth percentile of z-scores within an article.

Thus, LNHC would denote low novelty (LN) and high conventionality (HC), with all four combinations of LN and HN with LC and HC being possible. The authors observed that HNHC articles were twice as likely to be highly cited compared to the background rate, suggesting that novel combinations of ideas flavoring a body of conventional thought were a feature of impact. 

Key to the findings of Uzzi {\em et al.}~is the random graph model used, and its underlying assumptions. The citation switching algorithm used to generate expected values by substituting cited references with randomly selected references published in the same year is designed to preserve the number of publications, the number of references in each publication, and the year of publication of both publications and references. Importantly, disciplinary origin does not affect the probability that a reference is selected to replace another one. For example, a reference in quantum physics can be substituted, with equal probability, by a reference published in the same year but from the field of  quantum physics, quantum chemistry, classical literature, entomology, or anthropology. Such substitutions do not account for the disciplinary nature of scientific research and citation behavior \cite{wallace_lariviere_gingras_2012,moed_measuring_2010,klavans_research_2017,garfield_1979} very well. Accordingly, model misspecification is likely to arise on account of the simulated values not corresponding to the empirical data very well.

A follow-up study by Boyack and Klavans (2014) \cite{boyack_vs_uzzi_2014} explored the impact of discipline and journal effects on these definitions of conventionality and novelty.  While their study had some methodological differences in the use of Scopus data rather than WoS data, a smaller data set, and a $\chi^2$ calculation rather than Monte Carlo simulations to generate expected values of journal pairs, Boyack and Klavans noted strong effects from disciplines and journals. While they also reported the trend that HNHC is more probable in highly cited papers, they observed that ``only 64.4\%  of  243  WoS  subject  categories'' in the Uzzi {\em et al.} study met the criterion of having the highest probability of hit papers in the HNHC category.  Further, they observed that journals vary widely in terms of size and influence and that 20 journals accounted for 15.9\% of co-citations in their measurements. Lastly, they noted that three multidisciplinary journals accounted for 9.4\% of all atypical combinations. 

Despite different methods used to generate expected values, both of these key preceding studies measured co-citation frequencies across the scientific literature (using either WoS or Scopus) and normalized them without disciplinary constraints before subsequently analyzing disciplinary subsets. We hypothesized instead that modifying the normalization to constrain substitution references to be drawn only from the citation network being studied (the ``local network'') rather than all of WoS
(the ``global network'') would reduce model misspecification by limiting substitutions from references that were ectopic to these networks. Consequently, we used keyword searches of the scientific literature to construct exemplar citation networks themed around academic disciplines of interest: \emph{applied physics, immunology, and metabolism}. The cited references in these networks while predominantly aligned with the parent discipline (physics or life sciences in this case), also included articles from other disciplines. Within these disciplinary frameworks, we calculated observed and expected co-citation frequencies using a refined random graph model and an efficient Monte Carlo simulation algorithm.

Our analyses, using multiple techniques, provide substantial evidence that a constrained model where reference substitutions are limited to a local (disciplinary) network reduces model misspecification compared to the unconstrained model that uses the global network (WoS). Furthermore, re-analyses of these three semantically-themed citation networks  under the improved model reveals strikingly different trends. For example, while Uzzi {\em et al.} reported that highly cited articles are more likely than expected to be both HC and HN and that this trend largely held across all disciplines, we find that these trends vary with the discipline so that universal trends are not apparent. Specifically,  HC remains highly correlated with highly cited articles in the immunology and metabolism data sets but not with applied physics, and HN is highly correlated with highly cited articles in applied physics but not with immunology and metabolism.  Thus, disciplinary networks are different from each other, and trends that hold for the full WoS network do not hold for even large networks (such as metabolism).  Furthermore, we also found that the categories  demonstrating the highest percentage of highly cited articles (e.g., HC, HN, etc.) are not robust with respect to varying thresholds for high citation counts or for highly novel citation patterns. Overall, our study, although limited to three disciplinary networks, suggests that co-citation analysis that inadequately considers disciplinary differences may not be very useful at detecting universal features of impactful publications.

\section{Materials \& Methods}

 \subsection{Bibliographic data} We have previously developed ERNIE, an open source knowledge platform into which we parse the Web of Science (WoS) Core Collection \cite{Keserci371955}. WoS data stored in ERNIE spans the period 1900-2019 and consists of over 72 million publications. For this study, we generated an analytical data set from years 1985 to 2005 using data in ERNIE. The total number of publications in this data set was just over 25 million publications (25,134,073), which were then stratified by year of publication. For each of these years, we further restricted analysis to publications of type Article. Since WoS data also contains incomplete references or references that point at other indexes, we also considered only those references for which there were complete records~(Table \ref{tab:summary_data}). For example, WoS data for year 2005 contained 1,753,174 publications, which after restricting to type Article and considering only those references described above resulted in 916,573 publications, 6,095,594 unique references (set of references), and 17,167,347 total references (multiset of references). Given consistent trends in the data (Table \ref{tab:summary_data}), we analyzed the two boundary years (1985 and 2005) and the mid-point (1995). We also used the number of times each of these articles was cited in the first 8 years since publication as a measure of its impact.

\begin{table}[ht]
\caption{Summary of base WoS Analytical data set. Only publications of type Article with at least two references and references with complete publication data were selected for this data set. The number of unique publications of type Article, unique references (ur), total references (tr), and the ratio of total references to unique references increases monotonically with each year indicating that both the number of documents and citation activity increase over time. } 
\label{tab:summary_data}
\centering
\scalebox{0.7}{
\begin{tabular}{|r  cccc |}
  \hline
Year & Unique Publications & Unique References (ur) & Total References (tr) & tr/ur \\ 
  \hline
1985 & 391,860 & 2,266,584 & 5,588,861 & 2.47 \\ 
  \hline
1986 & 402,309 & 2,316,451 & 5,708,796 & 2.46 \\ 
1987 & 412,936 & 2,427,347 & 5,998,513 & 2.47 \\ 
1988 & 426,001 & 2,545,647 & 6,354,917 & 2.50 \\ 
1989 & 443,144 & 2,673,092 & 6,749,319 & 2.52 \\ 
1990 & 458,768 & 2,827,517 & 7,209,413 & 2.55 \\ 
1991 & 477,712 & 2,977,784 & 7,729,776 & 2.60 \\ 
1992 & 492,181 & 3,134,109 & 8,188,940 & 2.61 \\ 
1993 & 504,488 & 3,278,102 & 8,676,583 & 2.65 \\ 
1994 & 523,660 & 3,458,072 & 9,255,748 & 2.68 \\ 
  \hline
1995 & 537,160 & 3,680,616 & 9,875,421 & 2.68 \\ 
  \hline
1996 & 663,110 & 4,144,581 & 11,641,286 & 2.81 \\ 
1997 & 677,077 & 4,340,733 & 12,135,104 & 2.80 \\ 
1998 & 693,531 & 4,573,584 & 12,728,629 & 2.78 \\ 
1999 & 709,827 & 4,784,024 & 13,280,828 & 2.78 \\ 
2000 & 721,926 & 5,008,842 & 13,810,746 & 2.76 \\ 
2001 & 727,816 & 5,203,078 & 14,261,189 & 2.74 \\ 
2002 & 747,287 & 5,464,045 & 15,001,390 & 2.75 \\ 
2003 & 786,284 & 5,773,756 & 16,024,652 & 2.78 \\ 
2004 & 826,834 & 6,095,594 & 17,167,347 & 2.82 \\ 
   \hline
 2005 & 886,648 & 6,615,824 & 19,036,324 & 2.88 \\ 
 \hline
\end{tabular}}
\end{table}

We constructed three disciplinary data sets in areas of our interest based on the keyword searches: immunology,   metabolism, and   applied physics. For the first two, rooted in biomedical research, we searched Pubmed for the term `immunology' or `metabolism' in the years 1985, 1995, and 2005~(Table \ref{tab:disc-data sets}). Pubmed IDs (pmids) returned were matched to WoS IDs (wos\_ids) and used to retrieve relevant articles. For the applied physics data set, we directly searched traditional subject labels in WoS for `Physics, Applied'. While applied physics and immunology represent somewhat small networks (roughly 3-6\% of our analytical WoS datasets) over the three years examined, metabolism represents approximately 20-23\%, making them interesting and meaningful test cases. We also examined publications in the five major research areas in WoS: life sciences \& biomedicine, physical sciences, technology, social sciences, and arts \& humanities,  using the extended WoS subcategory classification of 153 sub-groups to categorize disciplinary composition of cited references in the data sets we studied.

\begin{table}[ht]
\caption{Disciplinary data sets. PubMed and WoS were searched for articles using search terms, `immunology', `metabolism', and `applied physics.' Counts of publications are shown for each of the three years analyzed and expressed in parentheses as a percentage of the total number of publications in our analytical WoS data set (Table \ref{tab:summary_data}) for that year. Note that Applied Physics and Immunology represent about 3-6\% of the publications in our analytical WoS datasets, but Metabolism occupies 20-23\%.}
\vspace{2 mm}
\label{tab:disc-data sets}
\centering
\begin{tabular}{|r r r r|}
  \hline
Year & Applied Physics & Immunology & Metabolism   \\ 
  \hline
1985 & 10,298 (2.7\%) & 21,606 (5.5\%) & 78,998 (20.2\%)  \\ 
1995 & 21,012 (3.9\%)  & 29,320 (5.5\%)  & 121,247 (22.6\%)   \\ 
2005 & 35,600 (4.0\%) & 37,296 (4.2\%) & 200,052 (22.6\%)    \\ 
 \hline
\end{tabular}
\vspace{-3mm}
\end{table}

\subsection{Monte Carlo simulations, normalization of observed frequencies, annotations, and `hit' papers}

We performed analyses on publications from 1985, 1995, and 2005. Building upon prior work \cite{uzzi_atypical_2013}, all ${n \choose 2}$ reference pairs were generated for each publication, where $n$ is the number of cited references in the publication. These reference pairs were then mapped to the journals they were published in using ISSN numbers as identifiers. Where multiple ISSN numbers exist for a journal, the most frequently used one in WoS was assigned to the journal. In addition, publications containing fewer than two references were discarded. Journal pair frequencies were summed across the data set to create observed frequencies $(F_{obs})$. 

For citation shuffling, we developed a performant citation switching algorithm, \emph{runtime enhanced permuting citation switcher (repcs)} \cite{GithubERNIE2019}, that randomly permuted citations within each disciplinary data set and within each year of publication: each citation within each article was switched within its permutation group in order to preserve the number of references from each publication year within each article. In so doing, the number of publications, the number of references in each data set, and the disciplinary composition of the references in each data set were preserved. Our approach is different from previous studies in these ways: (i) we sampled citations in proportion to their citation frequency (equivalently from a multiset rather than a set) in order to better reflect citation practice, (ii) we permitted a substitution to match the original reference in a publication when the random selection process dictated it rather than attempting to enforce that a different reference be substituted, and (iii) we introduced an error correction step to delete any publications that accumulated duplicate references during the substitution process.  As a benchmark, we used the citation switching algorithm of  \cite{uzzi_atypical_2013}, henceforth referred to  as \emph{umsj} as also done in \cite{boyack_vs_uzzi_2014}, using code kindly provided by the authors.  A single comparative analysis showed that while 10 simulations of the WoS 1985 data set (391,860 selected articles) completed in 2,186 hours using the \emph{umsj} algorithm, it completed in less than one hour using our implementation of the \emph{repcs}   algorithm on a Spark  cluster. We also tested \emph{repcs}  under comparable conditions to \emph{umsj} and estimated a runtime advantage of at least two orders of magnitude. This runtime advantage was significant enough that we chose to use the \emph{repcs}   algorithm in our study and generated expected values averaged over 1,000 simulations for improved coverage of every data set we analyzed. 

Using averaged results from 1,000 simulations for each data set studied, z-scores were calculated for each journal-pair using the formula $(F_{obs} - F_{exp})/\sigma$ where $F_{obs}$ is the observed frequency, $F_{exp}$ is the averaged simulated frequency, and $\sigma$ is the standard deviation of the simulated frequencies for a journal pair \cite{uzzi_atypical_2013}. As a result of these calculations, each publication becomes associated with a set of z-scores corresponding to the journal pairs derived from pairwise combinations of its cited references. Positional statistics of z-scores were calculated for each publication, which was then labeled according to conventionality and novelty: (i) HC if the median z-score exceeded the median of median z-scores for all publications and LC otherwise and (ii) HN if the tenth percentile of z-scores for a publication was less than zero and LN otherwise. We also analyzed the effect of defining high novelty using the first percentile of z-scores.

To consider the relationship between citation impact, conventionality, and novelty we calculated percentiles for the number of accumulated citations in the first 8 years since publication for each article we studied and stratified. We investigated multiple definitions of hit articles, with hits defined as the 1\%, 2\%, 5\%, and 10\% top-cited articles.  

\section{Results}

\subsection{Model Misspecification and the Attributes of Disciplinary Context} 
A source of misspecification arises from not accounting for disciplinary heterogeneity by treating all eligible references within WoS as equiprobable substituents when studying a disciplinary network. Under this model \cite{uzzi_atypical_2013}, the probability of selecting a reference from a discipline is identical to the proportion of the articles in WoS in that discipline for a given year.  If the global model accurately reflects citation practice, the expected proportion of references within papers published in a given discipline $D$ would be approximately equal to the proportion of references in  $D$, and conversely, the degree to which the proportion deviates from the expected value would reflect the extent of model misspecification. 

To study the disciplinary composition of references in our custom data sets, we first used the high level WoS classification of five major research areas: life sciences \& biomedicine, physical sciences, social sciences, technology, and arts \& humanities. The two largest of these research areas are physical sciences and life sciences \& biomedicine, which contribute on average approximately 35.1\% and 62.8\%, respectively, of the references in WoS over the three years of interest. Under the unconstrained model, we would expect close to 35\% of the references cited by the publications in any large network  to be drawn from the physical sciences and close to 63\% of the references to be drawn from  life sciences and biomedicine. Yet the empirical data present a very different  story: roughly 80\% of the references cited in physical sciences publications are from the physical sciences and 90\% of the references cited in life sciences \& biomedicine publications are from the life sciences \& biomedicine. In other words, the empirical data shows a strong tendency of publications to cite papers that are in the same major research area rather than in some other research area.  Thus, there is a strong bias towards citations that are {\em intra-network}. Our observations are in agreement with \cite{wallace_lariviere_gingras_2012} who found that, often, a majority of an article's citations are from the specialty of the article, even though that percentage varied among disciplines in the eight specialties they investigated (from approximately 39\% to 89\% for 2006). Furthermore, these findings argue that a discipline-indifferent random graph model would exhibit misspecification in deviating substantially from the empirical data, and supports the  concern about definitions of  innovation and conventionality that are based on deviation from expected values. 

\begin{figure}
\centering
\includegraphics[width=0.6\linewidth]{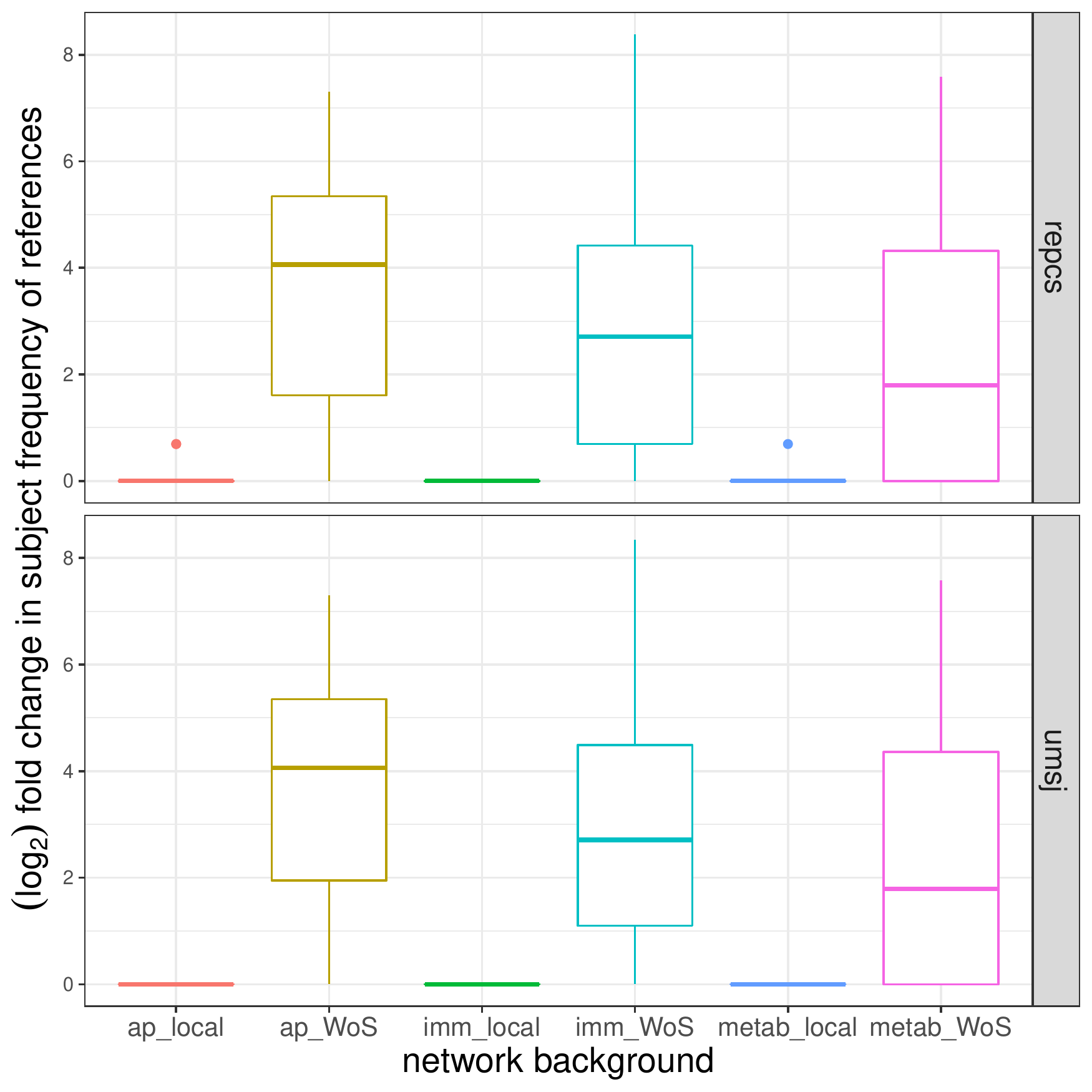}     
\caption{Citation shuffling using the local network preserves the disciplinary composition of references within networks, but using the global network does not.  Publications of type Article belonging to the three disciplinary networks (ap=applied physics, imm=immunology, and  metab=metabolism) were subject to a single shuffle of all their cited references using either the local network (i.e., the cited references in these networks, denoted  bg\_local) or the global network (i.e., references from all articles in WoS, denoted bg\_WoS) as the source of allowed substitutions, where ``bg'' indicates the disciplinary network. Citation shuffling was performed using either our algorithm (\emph{repcs}, top row)
 or that of Uzzi et al.~(\emph{umsj}, bottom row).
 The disciplinary composition of cited references before and after shuffling was measured as frequencies for each of 153 sub-disciplines (from the extended subject classification in WoS) and expressed as a fold difference between citation counts grouped by subject for original (o) and shuffled (s) references using the formula (fold\_difference = $ifelse(o > s, o/s, s/o)$) and rounded to the nearest integer. A fold difference of $1$ indicates that citation shuffling did not alter disciplinary composition. Data are shown for articles published in 1985. All eight boxplots are generated from 153 observations each. Null values were set to $1$. Note y-axis values: $log_2$} 
\label{fig:be}
\end{figure}

We also analyzed disciplinary composition at a deeper level using all 153 Subjects in the WoS extended classification and examining the consequences of  citation shuffling within a disciplinary set or all of the Web of Science.  References in publications belonging to these three data sets were summarized as a frequency distribution of 153 WoS Subjects as classes. A single shuffle of the references in the disciplinary data sets and in the corresponding WoS year slice was performed, using either the \emph{repcs}  or \emph{umsj} algorithms, after which subject frequencies were computed again. The fold difference in subject frequencies of references before and after shuffling was calculated for these groups using all 153 subject categories and summarized in the box plots in Fig \ref{fig:be}. As an example, the applied physics data set contained one reference labeled Genetics and Heredity, but after the shuffle (using the WoS background), acquired 1496 references labeled Genetics and Heredity. Similarly,  the metabolism data set  contained one reference labeled Philosophy, but after a single shuffle (again using the WoS background) it had 661 occurrences with this label. The data show convincingly that a publication's disciplinary composition of references in a network is preserved when citation shuffling is constrained to the network, but is significantly distorted when the WoS superset is used as a source of substitution. A second inference is that the two algorithms, \emph{repcs} and \emph{umsj}, have equivalent effects in this experiment (and so are only distinguishable for running time considerations).

\begin{table}[htbp]
\caption{Model misspecification is reduced by constraining substitutions to the local  disciplinary networks. We computed Kullback-Leibler (K-L) divergences between empirical and simulated journal pair frequencies using two different
background networks (local versus global) for each disciplinary network  (applied physics, immunology, and metabolism) for the years 1985, 1995, and 2005. K-L divergence was calculated using the R seewave package \cite{seewave2008}.  
For every disciplinary network, there is a smaller K-L divergence between simulated and observed data when using the local network (i.e., the disciplinary network) as compared to the global network (all of WoS). Put differently, model misspecification is reduced in the constrained model compared to the unconstrained model.}
\vspace{3mm}
\label{tab:kld}
\centering
\begin{tabular}{| ccccc |} 
  \hline
 Disciplinary Network & Year & Background Network & K-L Divergence & Ratio \\ 
  \hline
Applied Physics & 1985 & local & 1.21 &  \\ 
  & 1985 & global & 2.37 & 1.96 \\ 
  & 1995 & local & 0.86 &  \\ 
  & 1995 & global  & 2.37 & 2.77 \\ 
  & 2005 & local & 0.95 &  \\ 
  & 2005 & global  & 2.35 & 2.47 \\ 
    \hline
Immunology & 1985 & local & 0.75 &  \\ 
  & 1985 & global  & 1.68 & 2.24 \\ 
   & 1995 & local & 0.78 &  \\ 
   & 1995 & global  & 1.70 & 2.19 \\ 
   & 2005 & local & 0.73 &  \\ 
  & 2005 & global  & 1.92 & 2.63 \\ 
    \hline
Metabolism & 1985 & local & 1.11 &  \\ 
   & 1985 & global  & 2.24 & 2.02 \\ 
  & 1995 & local & 1.07 &  \\ 
  & 1995 & global  & 2.33 & 2.17 \\ 
     & 2005 & local & 1.19 &  \\ 
   & 2005 & global  & 2.60 & 2.18 \\ 
   \hline
\end{tabular}
\end{table}

 We then tested the conjecture that model misspecification would be reduced by constraining the substitutions to  disciplinary networks by examining the Kullback-Leibler (K-L) Divergence \cite{kullback_information_1951} between observed and predicted citation distributions, restricted to  the set of journals in a given disciplinary network.  
The results (Table \ref{tab:kld}) confirm this prediction: simulations under the constrained model 
(where the background network is the local disciplinary network) consistently have a lower K-L divergence compared to simulations under the unconstrained model  (where the background network is WoS).  Furthermore, the K-L divergence for the unconstrained model is generally twice as large as the K-L divergence for the constrained models, with ratios that range from 1.96 to 2.77, and are greater than 2.0 in eight out of nine cases. These results clearly demonstrate that constraining reference substitutions to the given local disciplinary network better fits the observed data, and hence reduces model misspecification.

\subsection{Calculation of Novelty and Conventionality using the constrained model}  
\begin{figure*}
\centering
\begin{tabular}{ccc}
\includegraphics[width=1.5in]{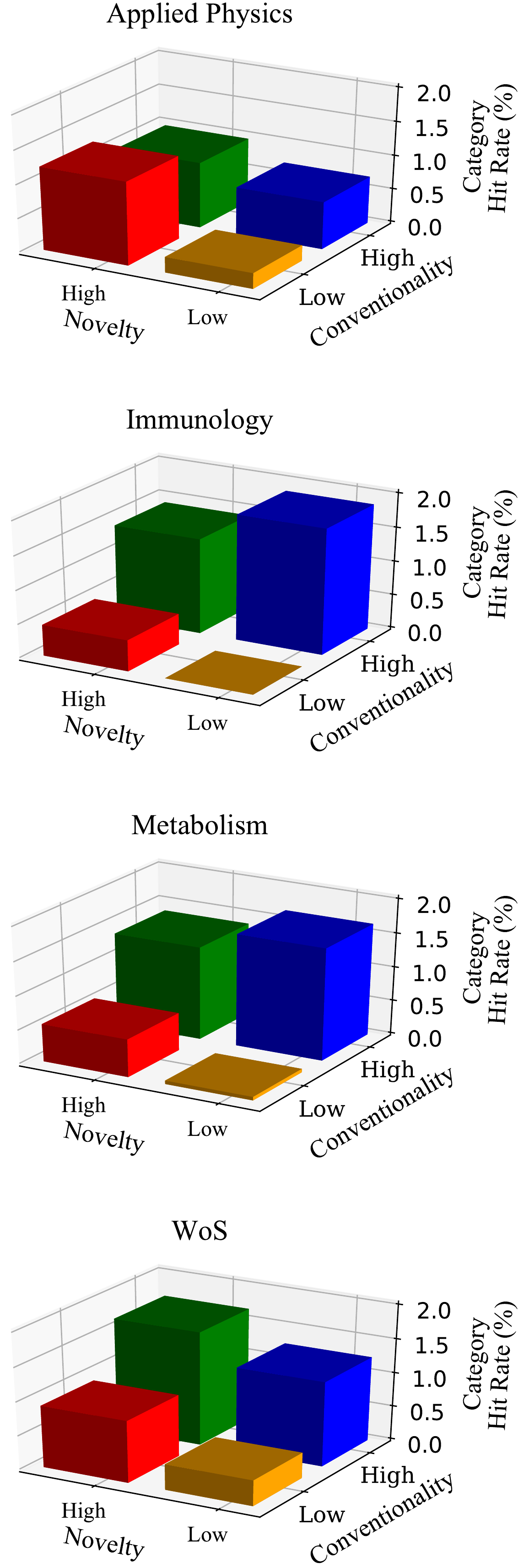} & \includegraphics[width=0.5in]{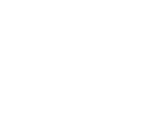} & \includegraphics[width=1.5in]{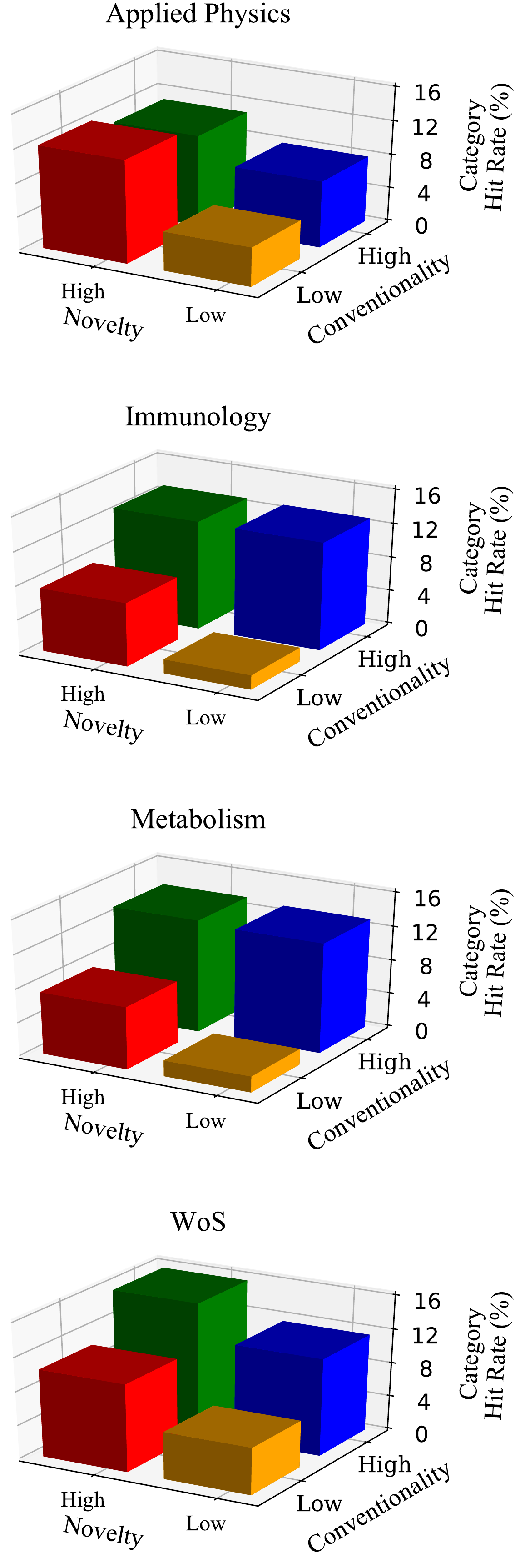} \\
(a) Top 1\% of cited articles & & (b) Top 10\% of cited articles \\
\end{tabular}
\caption{Effect of using the improved model on  categorical hit rates for Immunology, Applied Physics, and WoS for 1995. Panels (a) and (b) show hit rates for the LNLC, LNHC, HNLC, and HNHC categories for the applied physics, immunology, metabolism, and WoS data sets when hit articles are defined as the top 1\% and top 10\% of articles, respectively.  Novelty in both panels is defined at the 10th percentile of articles' z-score distributions. 
The results for the WoS data set also show that the highest hit rate is for the HNHC category.  Results for the three disciplinary networks all differ from the overall WoS results: the highest hit rates for the immunology and metabolism data sets  are in the LNHC category and the highest hit rate for the applied physics data sets are in the HNLC category. The number of data points in the applied physics, immunology, metabolism, and WoS data sets are 18,305, 21,917, 97,405, and 476,288, respectively.  
}
\label{fig:Fig2}
\end{figure*}

Since the constrained model better fits the observed data, we evaluated the distribution of highly cited articles  (i.e., ``hit articles'') in the four categories (HNHC, HNLC, LNHC, LNLC), for different thresholds for hit articles. Figure \ref{fig:Fig2}, Panels (a) and (b), compares hit rates for the four categories among the immunology, metabolism, applied physics, and WoS data sets for 1995, where the hit rate is defined as the number of hit articles in each category divided by the number of articles in the category. The calculation for the hit rates for the WoS data set (bottom row, Figure \ref{fig:Fig2}) mirrors Uzzi et al.'s results, whereby the largest hit rates were for the HNHC category, despite our methodological changes in sampling citations in proportion to their frequency.  However, the trends  for all three  disciplinary networks  are different from those for WoS. Specifically, the highest hit rates for the 1995 immunology and metabolism data sets are in the LNHC category for the top 1\% of cited articles (and tied between LNHC and HNHC for the top 10\%), and the highest hit rates for the 1995 applied physics data sets are in the HNLC category for both the top 1\% and top 10\% of all cited articles.  Thus, the category exhibiting the highest hit rate among highly cited papers depends on the specific disciplinary network and to some extent on the threshold for being highly cited.
 
Furthermore, the categories displaying the greatest hit rate vary to some extent with the year. For example, when the 10\% top-cited articles are deemed to be hits and novelty is defined at the 10th percentile of z-scores, the category with the highest hit rate in applied physics for 1995 is in HNLC (12.3\% versus 10.9\% for HNHC), while the hit rate for HNHC is greater than for HNLC in 1985 and 2005 (13.2\% versus 10.9\%, and 11.4\% versus 10.7\%, respectively).

We evaluated the statistical significance of the categorical hit rates using multiple methods.  Our first test was based on the null hypotheses that hits were distributed randomly among the four categories with uniform probability in proportion to the number of articles in each category. Rejecting the null hypothesis, using a Chi-Square Goodness of Fit test, supports a non-uniform dispersion of hits with some of the four categories being associated with higher or lower than expected expected hit rates. The null hypothesis was rejected at a $p<0.001$ in all cases in  Figure \ref{fig:Fig2}, with the exception of the immunology and applied physics data sets where hit articles are designated as the top 1\% of articles: valid tests were not possible in those instances due to too few expected hits. The null hypothesis was rejected with $p<0.001$ for all valid tests for all parameter settings, all data sets, and all years: hypotheses tests were valid in 73 of 96 instances. We conclude that it is likely that the distribution of hits among categories is not uniform and that, instead, hit rates vary among the categories in all disciplinary data sets. 

We also tested the explanatory power of each framework dimension by classifying articles as LN or HN and, separately, as LC or HC. We tested the null hypothesis that hits are distributed between LN and HN (LC and HC) in proportion to the total number of articles assigned to those categories.  That null hypothesis was rejected for the WoS data along both dimensions. Consistent with prior findings, hit articles were overrepresented in the HC category in every instance of WoS data at a $p<0.001$ and also overrepresented in the HN category at a $p<0.001$ in all but two cases: the p-values in those exceptions were $0.002$ and $0.007$.  Hits in the immunology and metabolism data were overrepresented in the HC category with the same statistical significance as for WoS. The relationship of novelty with hits in the immunology and metabolism data set differed dramatically from WoS, however, with statistically significant findings of hit articles being sometimes overrepresented in the LN category, and sometimes being underrepresented.  Consistent with WoS, hit articles in applied physics were positively related with HN with a statistical significance of at least $p<0.10$ in all 12 parameter sets, and at $p<0.05$ in 10 of 12 cases.  To the contrary, a strong positive relationship was found between LC and hit articles in applied physics in 5 of 12 instances with $p<0.10$. These results suggest that (1) both conventionality and novelty are strongly related to hits in WoS, (2) the conventionality dimension is strongly related with hits in immunology and metabolism and novelty is not, and (3) novelty is more strongly related with hits in applied physics than is conventionality. More generally, we find that the dimensions most strongly related with hit articles vary between disciplinary and broad data sets, and also among disciplines. 

We described concerns with model misspecification along two general dimensions: the background data set and sampling methodology for the random graph. The differences we found from prior research in terms of which categories demonstrated the highest hit rates were caused both by using disciplinary data sets and our sampling methodology, \textit{repcs},  through the article z-score distributions. 
When z-scores are shifted downward using one algorithm versus another, for example, then the former algorithm can result in an increased percentage of HN articles.
We therefore examined  the extent to which each of our methodological differences contributed to our observations.  
We found that z-scores changed sign more as a consequence of background network (local network or WoS) and much less as a consequence of sampling algorithm
 ({\textit{umsj} or \textit{repcs}). 
For example,  on the immunology data set, 28.6\% of the journal pairs changed signs with our sampling algorithm (\textit{repcs}) as the background network is changed from global (WoS) to local, and 
only 2.8\% of z-scores changed signs in the WoS data set depending on whether 
\textit{umsj} or \textit{repcs} was used. 

We conclude that the choice of background data sets is the source of a majority of differences we observed in the categories demonstrating the highest hit rates, although our sampling approach, most notably sampling from a multiset so as to reflect the observed frequencies of individual citations as well as their associated journals and disciplines, can also create material differences.  

\section{Discussion}

The principal difference between the two models we discuss is a single parameter--the set of references that can be used as substituents during the substitution process. The keyword search we use also has the advantage of selecting only relevant articles from multidisciplinary journals. However, it is important to note that the local networks we evaluated are not monodisciplinary, the references cited within exhibit disciplinary diversity. We provided several lines of evidence that showed that changing this one parameter from a global network to the local disciplinary network reduces model misspecification. Using the constrained model (which allows substitutions only within the local network)  instead of the unconstrained model (which allows substitutions in the WoS network) produces different trends in terms of conventionality and novelty, depending on the network and the parent discipline. In particular, when using the unconstrained model, highly cited papers were most likely to be in the HNHC category but this trend does not consistently hold when using the constrained model. Instead, we find that conventionality flavored with novelty is {\em not} generally a feature of impactful research. Further,  high ``novelty'' is not always indicative of impactful research.

More generally, these results show that the trends approaching universality in highly cited papers are not robust to changes in thresholds for defining high impact or high novelty articles, or with time, and may be the consequence of using a random model that has a poor fit to the observed data. On the other hand, while  the constrained model reduces model misspecification compared to the unconstrained model, this does not imply that the constrained model is reasonable nor that trends observed under the constrained model convincingly explain scientific practice. Indeed, there are significant  challenges in using random models to understand human behavior, of which citation practice is one example. As we note, \emph{vide supra}, under our conditions of analysis, the trends for all three disciplinary networks are different from those for WoS.

Our work has shown that the use of local networks enables simulations that are more consistent with research citation patterns. Further work might explore additional constraints on random assignment of citations to publications to better align benchmarks with citation practice. For example, proximity defined by co-author networks \cite{wallace_lariviere_gingras_2012} might be considered when defining probabilities for citation substitutions. Another interesting but challenging direction would be to find ways to distinguish intra-disciplinary from cross-disciplinary novelty. In this respect, the related work of \cite{wang_bias_2017} is insightful with its use of empirical data and observations made on novelty and quality, as well as dispersion and kinetics of accrued citations of articles classified as novel.

We note that journals are used as grouping units for articles in the three studies we discuss \cite{wang_bias_2017, uzzi_atypical_2013, boyack_vs_uzzi_2014} as well as this one. While we used keyword searches to identify sets of articles, we still relied on journal grouping to generate z-scores. 
Such a grouping, while appealing on account of relative simplicity, obscures measurements of novel pairings at the article level. Journals are also of limited use in representing individual fields, and repeating some of these studies using article clusters may be more informative \cite{traag_louvain_2019, klavans_which_2017}. Various factors contribute to citation counts \cite{peters_determinants_1994, vieira_citations_2010} and further study of these in the context of co-citation analysis may be of interest. We also acknowledge the limitations of using citation counts to identify impactful publications. Overall, evaluation in context \cite{hicks_bibliometrics:_2015} and further consideration of the disciplinary nature of the scientific enterprise is likely to result in improved models that yield further knowledge.

\section{Acknowledgments}

We thank two anonymous reviewers for helpful comments. We thank the authors of Uzzi et al. \cite{uzzi_atypical_2013} who kindly shared their Python code for citation shuffling. We are grateful to Kevin Boyack and Dick Klavans for constructively critical discussions.  We also thank Stephen Gallo and Scott Glisson for helpful suggestions. Research and development reported in this publication was partially supported by Federal funds from the National Institute on Drug Abuse, National Institutes of Health, US Department of Health and Human Services, under Contract Nos. HHSN271201700053C (N43DA-17-1216) and HHSN271201800040C (N44DA-18-1216). The content of this publication is solely the responsibility of the authors and does not necessarily represent the official views of the National Institutes of Health. TW receives funding from the Grainger Foundation.    

\section{Competing Interests}

The authors have no competing interests. Web of Science data leased from Clarivate Analytics was used in this study. Clarivate Analytics, had no role in conceptualization, experimental design, review of results, conclusions presented, and funding. Avon Davey's present affiliation is GlaxoSmithKline, Research Triangle Park, NC, USA. His contributions to this article were made while he was a full time employee of NET ESolutions Corporation.

\section{Data Availability} Access to the bibliographic data analyzed in this study requires a license from Clarivate Analytics. We have made supplementary data available on Mendeley Data at \\DOI: 10.17632/4n8ns8vzvz. Code generated for this study is freely available from our Github site \cite{GithubERNIE2019}.

\section{Author Contributions}
Conceptualization, GC, JB, SD, and TW; Methodology, AD, DK, GC, JB, SD, SL, and TW; Investigation, DL-H, GC, JB, and SD; Writing -Original Draft, GC, JB, TW; Writing- Review and Editing, AD, DK, DL-H, GC, JB, SD, SL, and  TW; Funding Acquisition, GC; Resources, DK and GC; Supervision, GC. Authors are listed in alphabetic order.

\bibliographystyle{plain}
\bibliography{arxiv2}
\end{document}